\documentclass{ocephys}

\usepackage{floatpag}

\newcommand{\pp}[2]{\frac{\partial #1}{\partial #2}}

\DeclareSIUnit{\cpkm}{cpkm}
\DeclareSIUnit{\days}{days}

\title{Assessing submesoscale sea surface height signals from the SWOT mission}

\author{Xihan Zhang}
\author{J\"orn Callies}
\affil{California Institute of Technology, Pasadena, CA}

\begin{document}
\nolinenumbers

\section*{Key points}

\begin{itemize}
  \item SWOT captures meso- and submesoscale turbulence whose spectral characteristics are broadly consistent with expectations.
  \item Compact cyclones have relative vorticities in excess of the planetary vorticity, challenging the usual interpretive framework for altimetry.
  \item The amplitude of a small-scale signal pervasive in the data is correlated in space and time with the heights of surface gravity waves.
\end{itemize}

\section*{Abstract}

The sea surface height (SSH) field measured by Surface Water and Ocean Topography (SWOT) mission's wide-swath altimeter is analyzed with a focus on submesoscale features. Along-track wavenumber spectra of SSH variance are estimated for the global ocean using the 1-day repeat period from March~26 to July~10, 2023. In regions with an energetic mesoscale eddy field, the spectra have a mesoscale plateau, a steep drop-off due to balanced submesoscale turbulence, and a much flatter power-law tail at small scales. These spectra are characterized by fitting a spectral model. For the balanced signal, this fit yields a power law exponent between $-4$ and $-6$ for most regions, broadly consistent with expectations and previous observations. The amplitude of the distinct small-scale signal, which typically dominates at wavelengths less than \qtyrange{30}{50}{\kilo\meter}, is strongly correlated in time and space with the height of surface gravity waves, suggesting aliased wave signals as the most likely source. A simple method is proposed to isolate the balanced signal in regions with negligible internal tides. Maps of the balanced signal in the Antarctic Circumpolar Current show compact cyclones with geostrophic relative vorticities frequently in excess of the local planetary vorticity, challenging the quasi-geostrophic framework commonly used to interpret altimetric data.

\section*{Plain language summary}

The recently launched Surface Water and Ocean Topography (SWOT) mission measures the elevation of the sea surface with unprecedented precision. The data returned from the early phase of this mission reveal four types of signals: the highs and lows associated with vortices, fronts, and filaments that are some \qtyrange{10}{100}{\kilo\meter} in size, the undulations associated with internal gravity waves excited by tidal flow over topography, a small-scale signal that appears to be linked to surface gravity waves, and variations in the time mean sea surface caused by the gravitational anomalies associated with seamounts, ridges, and other bathymetric features. The data open up new avenues to study turbulence in the upper ocean that is thought to be important in the exchange of heat, carbon, nutrients, and other tracers between the surface and interior ocean.

\section{Introduction}

Since the launch of TOPEX/POSEIDON in 1992, global satellite altimetry has greatly advanced our understanding of the ocean's large-scale circulation and mesoscale eddy field. Measurements of the sea surface height (SSH) relative to the geoid have produced accurate global tidal maps \parencite[e.g.,][]{le_provost_ocean_1995}, revealed a significant energy conversion from the external to the internal tide, exposed the long-range propagation of low-mode internal tides \parencite{ray_surface_1996}, provided a long-term record of sea level rise \parencite[e.g.,][]{cazenave_contemporary_2010}, and characterized the global field of mesoscale eddies, including their propagation and interaction with the large-scale circulation \parencite[e.g.,][]{wunsch_satellite_1998,chelton_global_2011,morrow_ocean_2023}. The noise level of order \qty{e2}{\centi\meter\squared\per\cpkm} limits the along-track resolution to about \qty{100}{\kilo\meter} wavelength, however, largely restricting nadir altimetry to the study of mesoscale and larger features \parencite[e.g.,][]{callies_expectations_2019}. Because ground tracks are spaced by order \qty{100}{\kilo\meter} and occupied on a repeat cycle of order \qty{10}{\days}, the maps produced by combining multiple passes and cycles have even coarser resolution and leave submesoscale turbulence inaccessible with nadir altimetry.

Numerical models and in situ observations have revealed that the upper ocean can host energetic submesoscale turbulence that is often associated with vigorous vertical exchange \parencite[e.g.,][]{mcwilliams_submesoscale_2016,taylor_submesoscale_2023}. As the horizontal grid spacing is reduced in numerical simulations, vertical velocities increase dramatically in magnitude as submesoscale fronts, filaments, and vortices are resolved \parencite{capet_mesoscale_1_2008}. This enhanced vertical flow has been suggested to play a leading-order role in the exchange of heat, carbon, and nutrients between the surface and interior ocean \parencite[e.g.,][]{levy_bringing_2012,mahadevan_impact_2016,balwada_submesoscale_2018,su_ocean_2018}, and it is thought to be important in structuring the marine ecosystem \parencite[e.g.,][]{levy_role_2018,freilich_diversity_2022}. Submesoscale dynamics may also drain energy from mesoscale eddies by a nonlinear energy transfer to small scales that is facilitated by frontogenesis \parencite[e.g.,][]{capet_mesoscale_3_2008,srinivasan_forward_2023} and an extraction of geostrophic kinetic energy by symmetric instabilities and slantwise convection \parencite[e.g.,][]{thomas_symmetric_2013}. Although much has been learned from numerical models and in situ observations, we are still lacking global observational constraints.

A primary goal of the Surface Water and Ocean Topography (SWOT) mission is therefore to characterize submesoscale dynamics globally \parencite{fu_observing_2008}. SWOT's Ka-band radar interferometer (KaRIn) provides SSH measurements with greatly improved precision \parencite{fu_surface_2024}. The measurements along two \qty[number-unit-product=-]{50}{\kilo\meter}-wide swaths provide two-dimensional maps directly, eliminating the need to interpolate in space and time to study submesoscale features \parencite{archer_wide-swath_2025}. This study utilizes the initial rapid-repeat phase of SWOT to assess the nature of submesoscale SSH signals captured by these new measurements.

Wavenumber spectra partition the variance of a signal across spatial scales, thus offering a convenient low-order statistical description of meso- and submesoscale velocity and SSH signals. Power laws are commonly observed, and the spectral slope (in log--log space) can be predicted by turbulence theory under---at times severe---idealizing assumptions~\parencite[e.g.,][]{charney_geostrophic_1971,blumen_uniform_1978,held_surface_1995}. In situ observations from the subtropical North Atlantic suggest that there are two regimes of submesoscale turbulence \parencite{callies_seasonality_2015}: an energetic regime with the kinetic energy (KE) spectrum following a $k^{-2}$ power law and a weak regime with the KE spectrum following a $k^{-3}$ power law. The energetic regime is typically observed in the winter mixed layer, whereas the weak regime is observed at the surface in summer and in the interior year-round. This suggests baroclinic instabilities in deep winter mixed layers are the primary mechanism energizing submesoscale turbulence \parencite{callies_role_2016}. Most of the KE injected by these instabilities at submesoscales subsequently cascades back to larger scales, a progression that can be seen in numerical models \parencite{sasaki_impact_2014} and nadir altimetry \parencite{lawrence_seasonality_2022}.
 
Wavenumber spectra have been studied extensively with nadir altimetry. Under the assumption of geostrophic balance, SSH variance spectra are related to the cross-track component of the KE spectrum by
\begin{equation}
  \langle |\hat{h}(k)|^{2} \rangle = \frac{f^{2}}{g^{2}k^{2}} \langle |\hat{v}(k)|^{2} \rangle,
  \label{geobalance}
\end{equation}
where $h$ is the SSH, $v$~is the cross-track velocity, $f$~is the planetary vorticity, $g$~is the gravitational acceleration, $k$~is the along-track wavenumber, the carets denote Fourier transforms, and the angle brackets denote an average over cycles. Thus, power laws in the KE spectrum can be translated to power laws in the SSH variance spectrum by subtracting~2 from the exponent. The strong submesoscale regime has a $k^{-4}$ SSH variance spectrum, while the weak submesoscale regime produces a $k^{-5}$ power law in SSH. With three years of TOPEX/POSEIDON data, \textcite{stammer_global_1997} found that SSH wavenumber spectra generally follow a $k^{-5}$ power law in extratropical regions. In contrast, \textcite{le_traon_altimeter_2008} argued that the spectra in western boundary current regions are consistent with $k^{-11/3}$ and therefore closer to the prediction of surface quasi-geostrophic rather than interior quasi-geostrophic turbulence theory. \textcite{xu_global_2011,xu_effects_2012} extended the analysis to near-global coverage with Jason-1/2 data and found spectral slopes that were often much smaller than those inferred from in situ observations, especially away from western boundary currents and the Antarctic Circumpolar Current. This inconsistency is likely the results of not distinguishing between geostrophic flows and internal tides, as internal tides dominating at small scales can drastically flatten SSH variance spectra~\parencite{richman_inferring_2012}. Carefully screening for internal tides, \textcite{lawrence_seasonality_2022} found spectral slopes generally consistent with in situ observations and the seasonality induced by mixed-layer instabilities. Satellite altimeters outside the TOPEX/POSEIDON and Jason series, including Saral/AltiKa and Sentinel-3, have also been used for SSH spectral analysis and its seasonality, giving generally consistent results \parencite[e.g.,][]{vergara_revised_2019}.

The SWOT spacecraft launched in December 2022. Before transitioning to its science orbit with a 21-day return period, it spent 3.5~months on a rapid-repeat orbit for calibration and validation, during which only a fraction of the ocean was observed, but the return period was reduced to 1~day. In this study, we focus on the data from this rapid-repeat phase both because it allows us to accurately remove small-scale geoid errors and because it gives us access to the rapid evolution of submesoscale turbulence. We use the level~2 low-rate data product (version~C). The SSH anomalies used throughout have the external, solid-earth, and polar tides removed as well as a dynamical atmosphere correction applied. We use the model-based wet-troposphere and sea state corrections, we apply the cross-over correction as supplied by the mission, and we retain the full internal-tide field. The SSH anomalies are referenced to the CNES/CLS 2022 mean sea surface \parencite{schaeffer_cnes_2023}.

\section{SSH variance spectra}
\label{results:globalspectra}
 
To assess the signals captured by SWOT, we calculate one-dimensional wavenumber spectra of SSH variance globally. We separate each pass into \qty[number-unit-product=-]{1000}{\kilo\meter}-long segments, overlapped by 50\% to better capture dynamically distinct regions along the track. For each segment, the KaRIn data consist of \qty[number-unit-product=-]{50}{\kilo\meter}-wide swaths on either side of the nadir, with a \qty[number-unit-product=-]{20}{\kilo\meter}-wide gap in between. The spatial posting of the globally available low-rate data is at \qty{2}{\kilo\meter}, making each swath 25~points wide. The power spectral density of SSH is estimated for each segment by applying a Hann window in the along-track direction, performing a discrete Fourier transform for each cross-track position, and averaging over the cross-track position as well as all cycles of the rapid-repeat phase (cycles \numrange{471}{577} from 2023-03-26 to 2023-07-10). Only cycles for which the segment has no missing data are used in the calculation. The same procedure (without the cross-track averaging, of course) is applied to the nadir data, obtained by a Jason-class altimeter aboard SWOT, which has a posting of \qty{6.86}{\kilo\meter}.

\begin{figure}
  \includegraphics[scale=0.61]{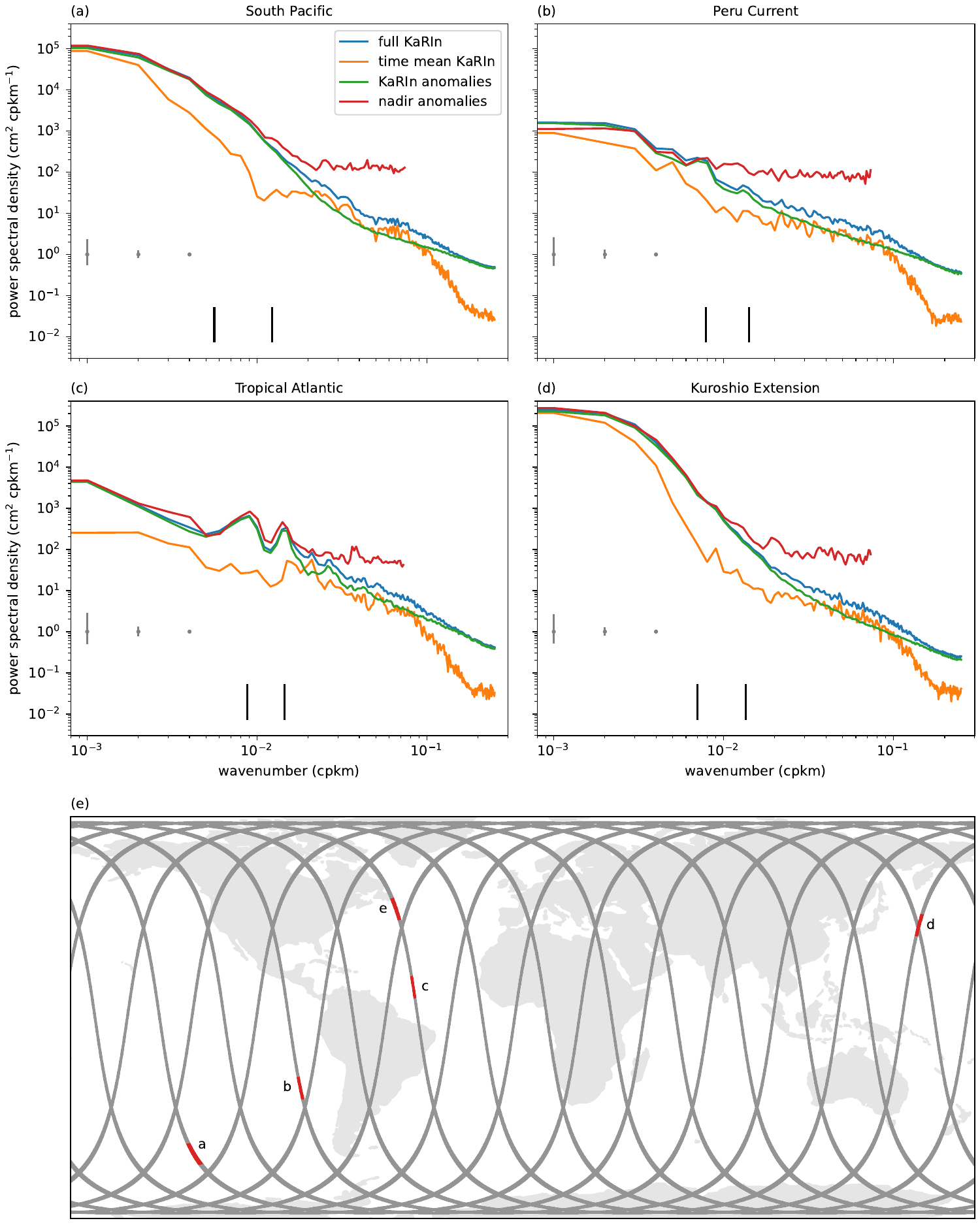}
  \caption{Overview of SSH variance spectra from SWOT's rapid-repeat phase. (a–d)~Examples of SSH variance spectra for the full KaRIn signal, the time mean KaRIn signal, the KaRIn anomalies, and the nadir anomalies. The vertical black lines indicate the wavenumbers of the M$_2$ internal tide for baroclinic modes 1 and~2. The vertical gray lines indicate 95\%~confidence intervals assuming that an independent sample is obtained every 10~cycles (left), every cycle (middle), and at every cross-track position (right). (e)~SWOT's rapid-repeat orbit and the locations of the five regions selected for the example spectra, among which a--d correspond to the respective panels here and region~e is shown in Fig.~\ref{fiteg}.}
  \label{swotpass}
\end{figure}

While all spectra are overall red, they vary substantially across the world ocean depending on the local dynamics (Fig.~\ref{swotpass}). We select four representative segments to illustrate the signals captured by SWOT: the South Pacific (Fig.~\ref{swotpass}a, centered on \ang{58}S, \ang{128}W), which crosses the Antarctic Circumpolar Current, the Peru Current (Fig.~\ref{swotpass}b, centered on \ang{89}W, \ang{28}S), the tropical Atlantic (Fig.~\ref{swotpass}c, centered on \ang{12}N, \ang{44}W), and the Kuroshio Extension region (Fig.~\ref{swotpass}d, centered on \ang{37}N, \ang{158}E).

\begin{figure}[t]
  \includegraphics[scale=0.61]{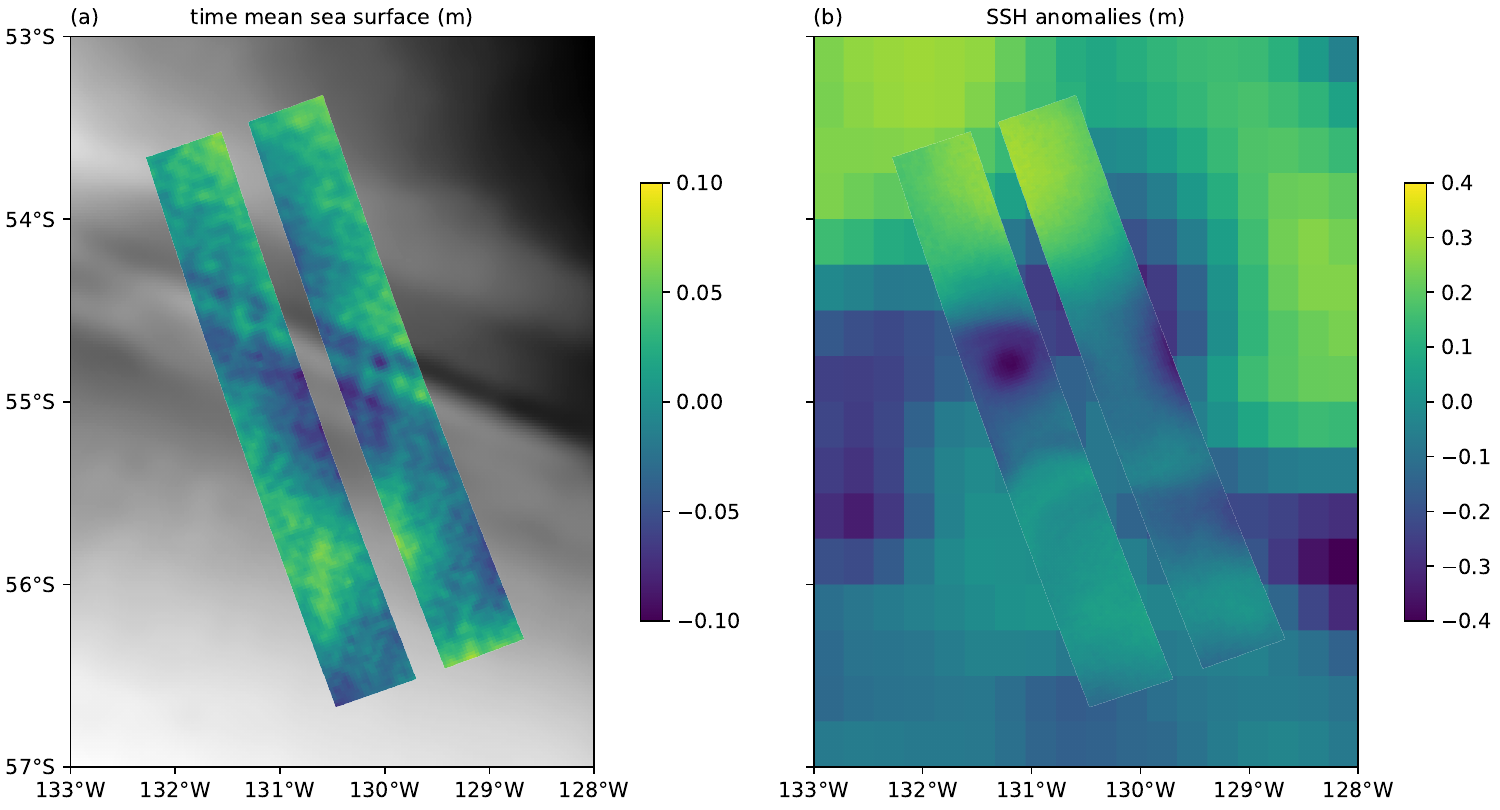}
  \caption{KaRIn's SSH signal in the South Pacific region. (a)~The time mean KaRIn signal (color shading) displaying small-scale geoid variations not removed by the prior geoid model. The variations line up with the independent ETOPO geoid model (gray shading). (b)~KaRIn's SSH anomalies (in-swath shading) and the DUACS product from nadir altimetry (background shading) on 24-04-2023 (SWOT cycle~500).}
  \label{sshmap}
\end{figure}

At wavelengths of about \qtyrange{10}{50}{\kilo\meter}, the mean sea surface contributes substantially to the KaRIn signal and produces a hump in the spectrum (Fig.~\ref{swotpass}a--d). The time mean height of the sea surface relative to a reference ellipsoid varies because of time mean geostrophic currents and geoid variations \parencite[e.g.,][]{wunsch_satellite_1998}. The mean sea surface from the CNES/CLS 2022 model is removed from the data, but small-scale variations of the mean sea surface were poorly constrained prior to SWOT. We therefore estimate these small-scale variations from a time mean over the SWOT data. The 3.5~months of the rapid-repeat phase are long enough to robustly estimate these small-scale variations, but the time mean also contains slowly varying mesoscale features. We therefore apply a two-dimensional Gaussian high-pass filter with a cutoff scale of \qty{100}{\kilo\meter} to the time mean signal before subtracting it from the SSH fields. We also remove a spatial mean for each segment to eliminate signals from the large-scale heating and cooling of the mixed layer \parencite{gill_theory_1973}.

The spectral level at wavelengths around \qtyrange{10}{50}{\kilo\meter} drops substantially if we remove the time mean, calculated from all available cycles (Fig.~\ref{swotpass}a--d). The variance of the time mean signal drops off precipitously at scales shorter than \qty{10}{\kilo\meter} wavelength until it hits a noise floor below \qty{e-1}{\centi\meter\squared\per\cpkm}. This drop-off is expected if this time mean signal is due to geoid variations caused by bathymetric features. Anomalies in the gravitational potential have a vertical decay scale equal to the horizontal wavenumber, so substantial reduction in the associated geoid variations is expected at wavelengths shorter than $2\pi$ times the water depth \parencite[e.g.,][]{smith_global_1997}. The inference that small-scale bathymetry dominates this time mean signal is also consistent with its patterns lining up with known bathymetric features, for example a ridge and canyon diagonally crossing the swaths in the South Pacific (Fig.~\ref{sshmap}a). The small-scale time mean sea surface can thus be used to infer small-scale bathymetry \parencite{yu_abyssal_2024}, although it should be kept in mind that the signal may, in principle, also contain an imprint from small-scale time mean currents. Still, we remove the time mean signal and focus our analysis on the anomalies. While topographic locking is known to occur for mesoscale eddies \parencite[e.g.,][]{holloway_systematic_1987,solodoch_formation_2021,siegelman_two-dimensional_2023}, more surface-trapped submesoscales should interact less strongly with bathymetry in the open ocean, if at all.

Spectral estimates follow chi-squared distributions if the underlying statistics are Gaussian \parencite[e.g.,][]{thompson_data_2014}. Because of the rapid repeat and dense spacing of pixels in the cross-track direction, the samples are not independent, which must be taken into account in the degrees of freedom of the chi-squared distribution used to calculate confidence intervals. Because this correlation in space and time depends on spatial scale, the number of degrees of freedom will be largest at small scales and smallest at large scales. To span the range of possible assumptions, we show three confidence intervals based on chi-squared statistics and three assumptions on the number of degrees of freedom: (i)~an independent sample is obtained every 10~cycles and all cross-track position are perfectly correlated, (ii)~an independent sample is obtained at every cycle and all cross-track position are perfectly correlated, and (iii)~an independent sample is obtained at every cycle and every cross-track position.

The power spectra of SSH anomalies in the South Pacific (Fig.~\ref{swotpass}a) and the Kuroshio Extension region (Fig.~\ref{swotpass}d) have a similar structure. They have a plateau at large scales and a mesoscale ($\sim$\qty{300}{\kilo\meter} wavelength) transition to a steep power-law drop-off at submesoscales. At \qtyrange{30}{50}{\kilo\meter} wavelengths, the spectra flatten out again and follow a different power law down to the Nyquist wavelength of \qty{4}{\kilo\meter}. We refer to the meso- and submesoscale signal above this transition to a flatter spectrum, i.e., at wavelengths greater than \qty{50}{\kilo\meter}, as the ``balanced signal.'' At mesoscales, the signal is expected to be geostrophically balanced. Nonlinear terms can enter the balance at submesoscales (see below), but this signal appears to remain balanced as opposed to the internal-wave signal seen elsewhere. We refer to the signal below the transition to a flatter spectrum, i.e., at wavelengths below \qty{30}{\kilo\meter}, as the ``small-scale signal,'' and we investigate its origin in Section~\ref{results:smallscale}. Part of the balanced signal is captured by the nadir data, but white noise with a spectral level of about \qty{e2}{\centi\meter\squared\per\cpkm} obscures any signals below about \qty{100}{\kilo\meter} wavelength.

Away from major current regions, internal tides also make a leading-order contribution to the spectra (Fig.~\ref{swotpass}b,c).\footnote{Tides are aliased in time even in the rapid-repeat phase. The alias period of the M$_2$ tide, for example, is \qty{12.4}{\days}. No attempt is made here to use temporal information to isolate the tidal signal.} Peaks in the spectrum can be made out at the wavelengths corresponding to low-mode semi-diurnal tides. The modal wavenumbers are calculated as $k_n = \sqrt{\omega^2 - f^2}/c_n$, where $c_n$ are the modal gravity wave speeds estimated from the ECCO~v4 climatology \parencite{forget_ecco_2015}. Part of this signal is captured by nadir altimeters; this signal is well-known \parencite[e.g.,][]{richman_inferring_2012,ray_m2_2016} and has been used to make maps of the low-mode internal tides globally \parencite[e.g.,][]{dushaw_empirical_2015,zhao_global_2016,zaron_baroclinic_2019}. In our examples, internal tides dominate over any submesoscale balanced signals in the Peru Current region (Fig.~\ref{swotpass}b), making it difficult to characterize submesoscale turbulence there, and it produces prominent peaks in the tropical Atlantic (Fig.~\ref{swotpass}c), where mesoscale signals give way to equatorial waves and jets. In all cases, there is a small-scale signal with a similar power law.

\begin{figure}[t]
  \includegraphics[width = \textwidth]{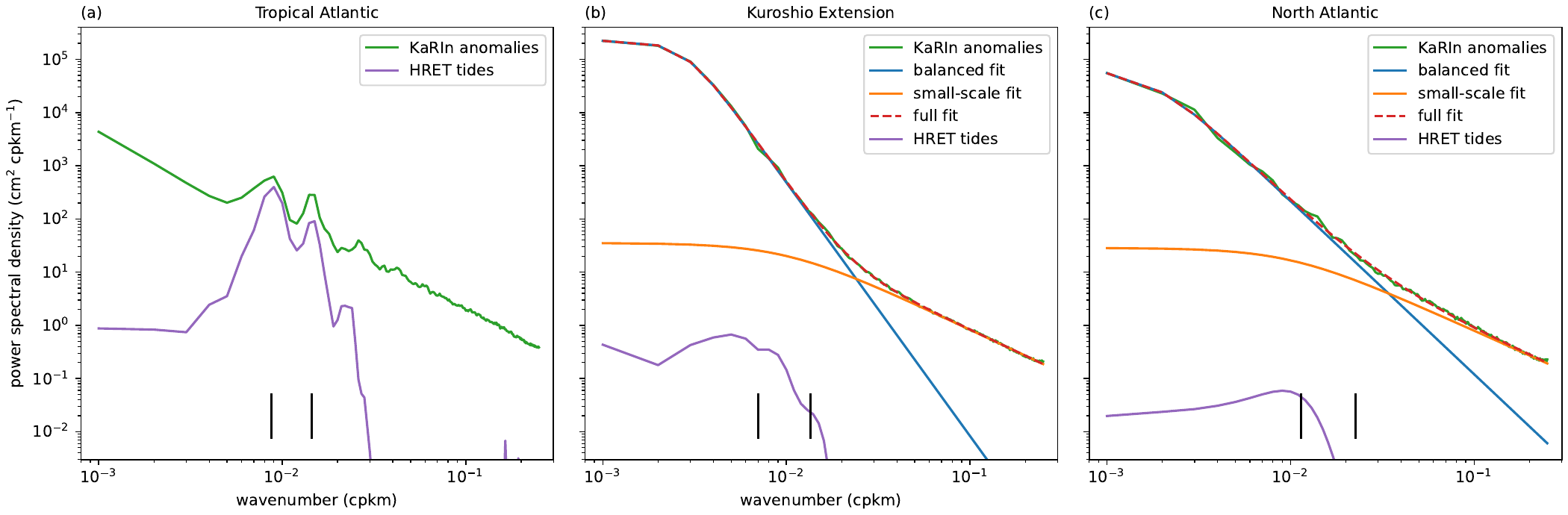}
  \caption{Tidal signatures and model fits to example SSH variance spectra. (a)~Spectrum of SSH anomalies and the HRET estimate of the internal-tidal signal in the tropical Atlantic region. (b)~Spectrum of SSH anomalies in the Kuroshio Extension region. (c)~Spectrum of SSH anomalies for a subpolar North Atlantic segment (location~e in Fig.~\ref{swotpass}e, centered on \ang{43}N, \ang{51}W). The spectra in (b) and~(c) have negligible tidal signals as estimated by HRET, and the fits of the model~\eqref{slopefiteq} to the balanced and small-scale components of the signal are shown. Panel~(b) shows a typical fit, whereas (c) shows an outlier with an anomalously flat balanced spectrum.}
  \label{fiteg}
\end{figure}

\begin{figure}
  \vspace{-1in}
  \includegraphics[scale=0.61]{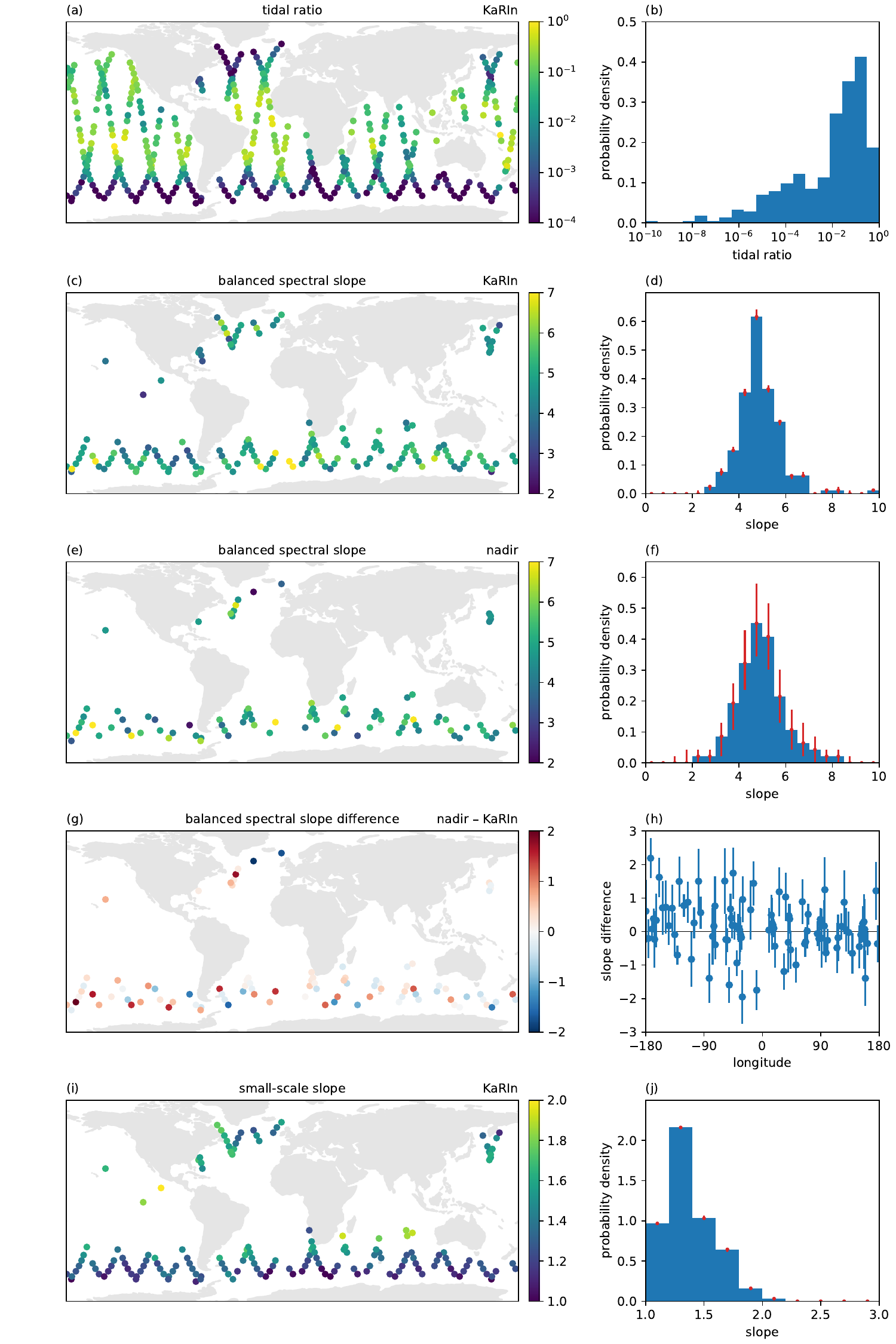}
  \caption{Spectral properties of the SSH signal. (a,b)~Map and histogram of the tidal ratio, the maximum of the ratio between the variance spectra of the HRET estimate for internal tides and the full KaRIn anomalies. (c,d)~Map and histogram of the slope~$s_\mathrm{b}$ of the balanced signal estimated from KaRIn data. (e,f)~Map and histogram of the slope~$s_\mathrm{b}$ of the balanced signal estimated from nadir data. (g,h)~Map and scatter plot of the difference between these two estimates of~$s_\mathrm{b}$. (i,j)~Map and histogram of the small-scale slope~$s_\mathrm{s}$ in KaRIn data.}
  \label{slopes}
  \thisfloatpagestyle{empty}
\end{figure}

We use the high-resolution empirical tide \parencite[HRET;][]{zaron_baroclinic_2019} estimate to screen all segments for possible contributions of low-mode internal tides to the KaRIn signal. This estimate is produced using harmonic fits to the long record of nadir data. As such, it may substantially underestimate tidal amplitudes in regions where Doppler shifts and the propagation in a time-dependent medium broaden the tidal lines \parencite[e.g.,][]{rainville_propagation_2006}, so we take it as a lower bound for the contribution to the signal by internal tides. In the tropical Atlantic, for example, the HRET signal clearly captures the prominent peaks due to the first two baroclinic modes of the semi-diurnal tide (Fig.~\ref{fiteg}a). In the Kuroshio Extension region, on the other hand, HRET explains only a tiny fraction of the full signal (Fig.~\ref{fiteg}b). We consider regions in which the HRET signal makes up more than 1\% of the spectral density at any wavenumber as significantly affected by internal tides and exclude these regions from the analysis of the submesoscale spectral slope that follows. This removes about 70\% of all segments, leaving only regions with a strong mesoscale eddy field \parencite[Fig.~\ref{slopes}a; cf.,][]{lawrence_seasonality_2022}.

To characterize the spectra and estimate the submesoscale spectral slope for the remaining segments, we fit to their SSH variance spectra a model that consist of two parts:\footnote{Throughout the paper, $k$ denotes an angular wavenumber, related to the wavelength~$\ell$ by $k = 2\pi/\ell$. The spectra are one-sided spectral densities with respect to this angular wavenumber, i.e., they are normalized such that $\int_0^\infty P(k) \, \d k$ equals the variance of the signal. As is conventional, however, we use the ordinary wavenumber $k/2\pi = \ell^{-1}$ (in cycles per kilometer, \unit{\cpkm}) in all plots to make it easier to read off the corresponding wavelength. The spectral density is multiplied by $2\pi$ to ensure a consistent normalization in the plots.}
\begin{equation}
  P(k; \theta) = \frac{A_\mathrm{b} (\lambda_\mathrm{b} k)^{p_\mathrm{b}}}{1 + (\lambda_\mathrm{b} k)^{s_\mathrm{b} + p_\mathrm{b}}} + \frac{A_\mathrm{s}}{[1 + (\lambda_\mathrm{s} k)^2]^{s_\mathrm{s}/2}},
  \label{slopefiteq}
\end{equation}
where $\theta = (A_\mathrm{b}, \lambda_\mathrm{b}, s_\mathrm{b}, p_\mathrm{b}, A_\mathrm{s}, \lambda_\mathrm{s}, s_\mathrm{s})$ is the collection of parameters to be estimated from the fit. The first term represents the balanced signal. The overall spectral level is set by~$A_\mathrm{b}$, and the model transitions from a power law $k^{p_\mathrm{b}}$ at low wavenumbers to a power law $k^{-s_\mathrm{b}}$ at high wavenumbers. The transition occurs at $k \sim \lambda_\mathrm{b}^{-1}$, i.e., a wavelength of $2\pi \lambda_\mathrm{b}$. The spectral slope~$s_\mathrm{b}$ at high (submesoscale) wavenumbers is the main target of this estimation. The second term represents the small-scale signal. It is captured with a Mat\'ern model and characterized by the spectral level~$A_\mathrm{s}$, the high-wavenumber spectral slope~$s_\mathrm{s}$, and the transition scale~$\lambda_\mathrm{s}$. The parameters are determined using a least-squares fit with a $k^{-1}$ weight to reduce the impact of the better-sampled small scales on the fit \parencite[cf.,][]{lawrence_seasonality_2022}. The transition scale~$\lambda_\mathrm{s}$ is poorly constrained by the data and set to $\qty{100}{\kilo\meter} / 2\pi$ throughout. For the segment in the Kuroshio Extension region, this fit captures the balanced and small-scale signals as well as the transition between the two (Fig.~\ref{fiteg}b). The balanced spectral slope is \num{4.77+-0.007} (2$\sigma$ uncertainty, where $\sigma$ is the standard error of the least squares fit)---corresponding to \num{2.77+-0.007} for KE spectrum, assuming geostrophic balance holds. The small-scale signal has a much smaller spectral slope of \num{1.63+-0.001} and dominates at wavelengths below \qty{40}{\kilo\meter}.
 
We fit the KaRIn spectra with the model~\eqref{slopefiteq} for all segments with negligible contribution from internal tides (Fig.~\ref{slopes}c,d). We further discard segments for which the uncertainty (2$\sigma$) for the balanced slope~$s_\mathrm{b}$ is greater than $1.0$. The histogram of the estimated balanced spectral slopes peaks near $s_\mathrm{b} = 5$, and 77\% of the slope estimates fall between 4 and~6. This steep drop-off of the balanced signal is expected \parencite{callies_expectations_2019}, consistent with previous analysis of nadir data \parencite[e.g.,][]{lawrence_seasonality_2022}, and broadly consistent with numerical simulations \parencite[e.g.,][]{capet_mesoscale_1_2008,torres_partitioning_2018} and KE spectra from in situ observations \parencite[e.g.,][]{callies_interpreting_2013,shcherbina_statistics_2013,callies_seasonality_2015,rocha_mesoscale_2016}. The spectra are steeper, on average, than the $s_\mathrm{b} = 4$ expected for energetic submesoscale turbulence. This may be the result of the rapid-repeat phase falling into spring to early summer in the northern hemisphere, when mixed layers have shoaled and mixed-layer instabilities have ceased to energize submesoscale turbulence, and into fall to early winter in the southern hemisphere, when mixed layers have not yet deepened for mixed-layer instabilities to have substantially energized submesoscale turbulence. A better constraint of the seasonality, however, should be sought using the science phase data. It should also be kept in mind that the model~\eqref{slopefiteq} is entirely empirical, and the fit to the observed spectrum is not always as good as for the Kuroshio Extension region (Fig.~\ref{fiteg}b). Many of the spectra with very low~$s_\mathrm{b}$, for example, a segment in the subpolar North Atlantic (Fig.~\ref{fiteg}c), lack a mesoscale plateau and a clearly distinct submesoscale power law.

For comparison, we also fit the nadir data with a spectral model. We use the same model for the balanced signal and add a white noise with spectral level~$N$:
\begin{equation}
  P(k; \vec{\theta}) = \frac{A_\mathrm{b} (\lambda_\mathrm{b} k)^{p_\mathrm{b}}}{1 + (\lambda_\mathrm{b} k)^{s_\mathrm{b} + p_\mathrm{b}}} + N,
  \label{slopefiteqnadir}
\end{equation}
where the parameter vector is now $\vec{\theta} = (A_\mathrm{b}, \lambda_\mathrm{b}, s_\mathrm{b}, p_\mathrm{b}, N)$. We calculate the nadir spectra using the same cycles as used for the KaRIn spectra, except that we discard additional cycles when gaps are present in the nadir data, so the number of cycles used for the nadir spectra is the same as for KaRIn or smaller. Together with the lack of cross-track averaging, this leads to higher uncertainties for the nadir spectra. The main reason for more poorly constrained spectral slopes~$s_\mathrm{b}$ (Fig.~\ref{slopes}e), however, is the elevated noise level (Fig.~\ref{swotpass}). There are fewer segments with a reliable slope estimate (2$\sigma$~uncertainty less than \num{1.0}) than for KaRIn. As for the KaRIn data, the histogram of estimated slopes peaks around $s_\mathrm{b} = 5$, but the nadir histogram is somewhat broader and more uncertain. The differences in the slope estimated from nadir and KaRIn data mostly fall within \num{+-1} and are scattered around~0, indicating consistency between the two data sets and general robustness of the approach. More importantly, KaRIn data constrain the slopes substantially better and allow for estimates in more regions than has been possible with nadir data.

\section{Small-scale signal}
\label{results:smallscale}

In contrast to the balanced signal, the small-scale signal has an origin that is not immediately apparent. The internal-wave continuum \parencite{garrett_space-time_1972} is one contender. Its signature in the SSH variance spectrum is expected to have a $k^{-2}$ roll-off, which is much gentler than that of the balanced flow and suggests that the internal-wave continuum should dominate strongly at sufficiently small scales \parencite{callies_expectations_2019}. While there are uncertainties in the prediction of the spectral level of the continuum because different assumptions can be made in translating the predictions of the interior wave amplitudes \parencite{garrett_space-time_1975,munk_internal_1981} to their SSH signature \parencite{samelson_models_2024}, two predictions appear robust. First, the expected spectral roll-off of $k^{-2}$ is not affected by uncertainties in the spectral level. Second, the SSH amplitude should vary seasonally with the near-surface stratification. The SSH signature of an internal-wave mode is governed by the seasonally modulated vertical structure of the mode, leading to enhanced SSH signals in summer and suppressed signals in winter \parencite{rocha_seasonality_2016,lahaye_sea_2019}. We show below that neither of these predictions is consistent with the data. The variations of the small-scale signal's amplitude in space and time suggests that they are linked to surface gravity waves instead.

The slope of the small-scale signals is estimated as part of the fit described in the previous section and is substantially smaller than $s_\mathrm{s} = 2$ almost everywhere (Fig.~\ref{slopes}i). The histogram of slopes peaks at the \numrange{1.2}{1.4}~bin (Fig.~\ref{slopes}j). Neither these small values nor their geographic variation is consistent with the internal-wave continuum.

To quantify the strength of the small-scale signal, we integrate the power spectral density of the SSH anomalies from KaRIn over the last octave (\qtyrange{4}{8}{\kilo\meter} wavelengths) and take the square root. We refer to this quantity as the ``small-scale amplitude,'' although we note that there is more variance of the small-scale signal above the \qty{8}{\kilo\meter} wavelength (Fig.~\ref{swotpass}a--d).

\begin{figure}[t]
  \includegraphics[scale=0.61]{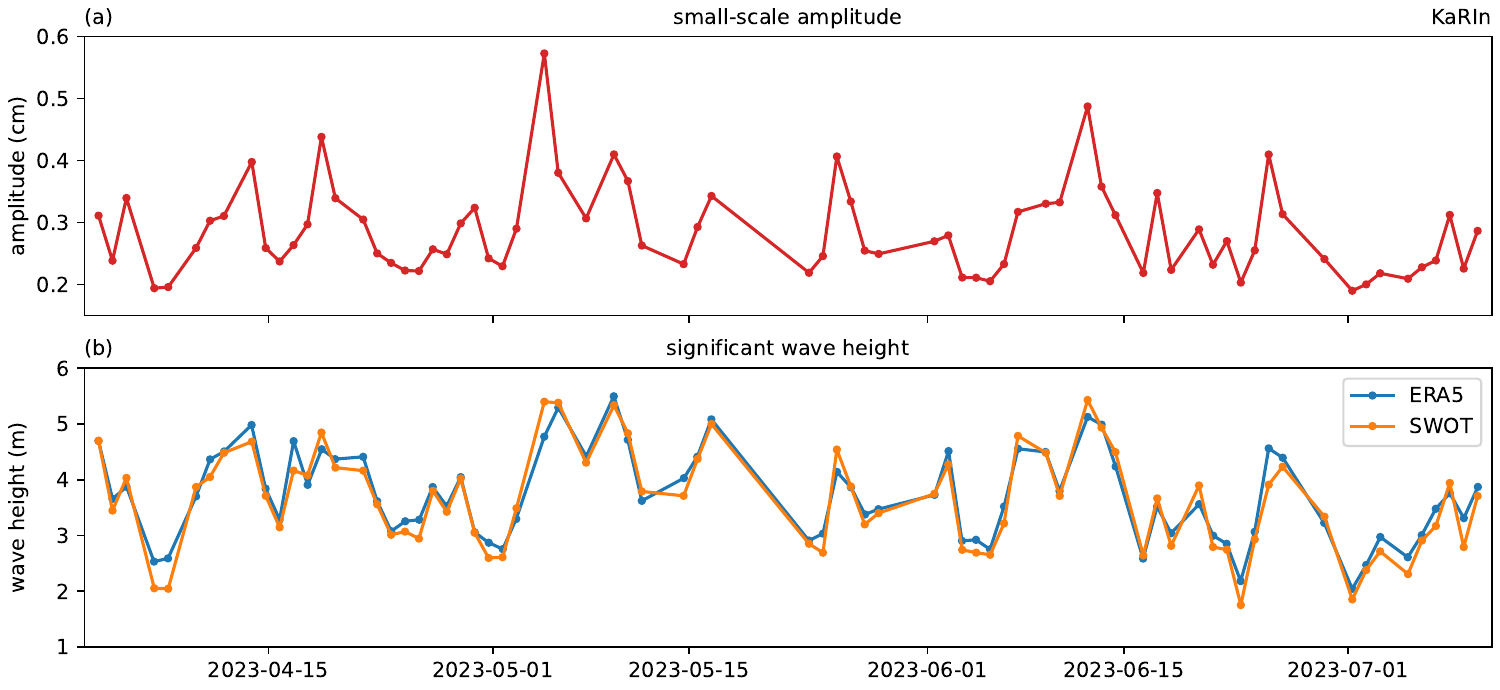}
  \caption{Correlations between the amplitude of the small-scale signal in KaRIn data and the significant wave height in the South Pacific region. (a)~Time series of the small-scale amplitude in the KaRIn data. (b)~Time series of the significant wave heights from ERA5 and as estimated from SWOT data.}
  \label{wavets}
\end{figure}

We first assess the temporal variability of the small-scale signal in the South Pacific example region (Figs.~\ref{swotpass}a, \ref{sshmap}). The small-scale amplitude in this region varies by a factor of three and fluctuates substantially from cycle to cycle (Fig.~\ref{wavets}). These variations are strongly correlated with the height of surface gravity waves in the region. We obtain estimates for the wave heights from ERA5 \parencite{hersbach_era5_2020} and from the SWOT product itself, which is based on the KaRIn backscatter and an empirical relation to the wave height. These two estimates of the significant wave height are very similar to one another and strongly correlated with the variations in the small-scale amplitude. The correlation coefficients with the wave heights from ERA5 and SWOT are \numlist{0.72;0.84}, respectively, suggesting the wave height controls the spectral level of the small-scale signal in this region.

\begin{figure}[t]
  \includegraphics[scale=0.61]{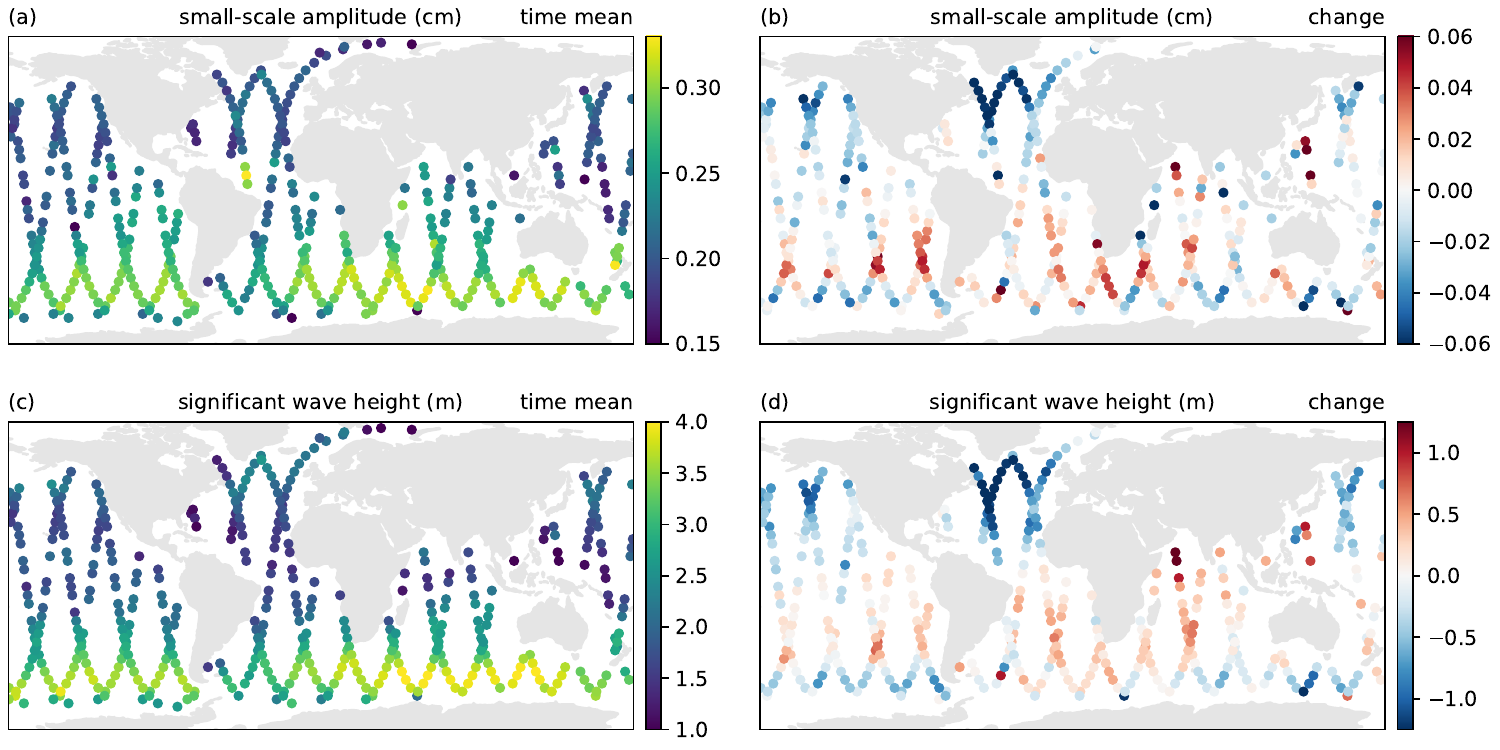}
  \caption{Spatial and seasonal patterns of the small-scale amplitude of the KaRIn signal and their correspondence to significant wave heights. (a)~Time mean small-scale amplitude of the KaRIn signal. (b)~Change in the small-scale amplitude over the course of the rapid-repeat phase. (c)~Time mean significant wave height. (d)~Change in the significant wave height over the course of the rapid-repeat phase. Change is defined as the time mean over the second half of the rapid-repeat phase minus the mean over the first half. Significant wave heights are estimated from SWOT data. ERA5 has a similar pattern (not shown).}
  \label{wavemaps}
\end{figure}

This correlation between surface gravity wave heights and the small-scale amplitude of the KaRIn signal applies in space and time globally (Fig.~\ref{wavemaps}). The geography of the time mean small-scale amplitude is remarkably similar to that of the significant wave height (Fig.~\ref{wavemaps}a,c). Both are much larger in the southern hemisphere than the northern hemisphere, with a strong enhancement in the South Indian Ocean. Both are suppressed at low latitudes and again regionally enhanced under the northern-hemisphere storm tracks, albeit much less strongly than in the Southern Ocean. The time mean small-scale amplitude and the significant wave height estimated from SWOT have a correlation coefficient of \num{0.85}. In addition, the seasonal trend over the 3.5-month rapid-repeat phase, computed as the average over the second half minus the first half, also has a similar geographic pattern for the two variables (Fig.~\ref{wavemaps}b,d). In the southern hemisphere, both variables pick up at midlatitudes and decrease somewhat at high latitudes as the storm track moves north during the transition from austral fall to winter. In the northern hemisphere, both variables decrease in the extratropics almost everywhere, with a particularly marked decrease in the North Atlantic. The striking similarity in these patterns suggests that wave heights control KaRIn's small-scale signal globally.

The geography and seasonal trends of the small-scale amplitude (Fig.~\ref{wavemaps}) are inconsistent with the interpretation of the small-scale signal as the internal-wave continuum. One would expect a smaller hemispheric asymmetry \parencite{samelson_models_2024} and the opposite seasonal trends from what is observed. During the transition to boreal summer, the increase in upper-ocean stratification should amplify the surface expression of internal waves in the northern hemisphere. During the transition to austral winter, the erosion of the upper-ocean stratification should suppress the surface expression of internal waves in the southern hemisphere. KaRIn's small-scale signal is clearly inconsistent with this expectation.

In summary, the amplitude of KaRIn's small-scale signal appears to be modulated by the height of surface gravity waves. Pre-launch estimates suggested a modulation of KaRIn's noise level by wave heights, and the expected noise levels were a few centimeters for typical wave conditions \parencite{peral_impact_2015}. How the waves might alias into the scales captured by the \qty{2}{\kilo\meter} low-rate product and give rise to the universally observed power-law spectrum, however, remains insufficiently understood.

It is also possible that other signals that are correlated in space and time with surface gravity waves contribute to KaRIn's small-scale signal. Near-inertial waves are also forced by the winds and therefore follow a similar pattern \parencite[e.g.,][]{alford_near-inertial_2016}. They can have large-amplitude velocities after the passage of a storm. While they have no leading-order expression in surface height, interaction with mesoscale eddies engenders spatial variations \parencite[e.g.,][]{thomas_near-inertial_2017,conn_interpreting_2024} that do produce a surface height signal. Near-inertial waves are substantially stronger in the North Pacific than the North Atlantic \parencite{yu_surface_2019}, however, so their geography does not appear to match that of KaRIn's small-scale amplitude as well as that of surface gravity waves does. More work is needed to better understand the potential signals of near-inertial waves in surface height.

\section{Maps of the balanced signal}
\label{results:balanced}

Given that the small-scale signal is clearly distinct from the balanced signal at larger scales and that it appears to be linked to surface gravity waves, we here attempt to isolate the balanced signal to study its patterns and evolution in maps. We use the South Pacific region as an example (Fig.~\ref{sshmap}). It is in the Antarctic Circumpolar Current and has a strong balanced signal as well as negligible low-mode internal tides. The method described below only removes the small-scale signal, so it can be applied everywhere the tidal signal is negligible, although the approach could be modified to allow for tidal signal as well.

We use a simple statistical approach to distinguish the balanced and small scale signals. We use ``balanced'' purely as a label for the meso- and submesoscale signal showing a steep spectral roll-off and emphasize that the extraction method that follows makes no assumptions on dynamical balances. We assume the observed signal $\vec{y}$, the pixel map stacked into a vector, to have two components: a balanced signal $\vec{x}$ and a small-scale signal $\vec{\eta}$:
\begin{equation}
  \vec{y} = \vec{x} + \vec{\eta}.
\end{equation}
We assume both of them to have Gaussian statistics, $\vec{x} \sim \mathcal{N}(\vec{0}, \hat{\mat{C}})$ and $\vec{\eta} \sim \mathcal{N}(\vec{0}, \mat{N})$, both with zero mean and prior covariance matrices $\hat{\mat{C}}$ and~$\mat{N}$. According to Bayes's theorem, the balanced signal given the observations~$\vec{y}$ is also Gaussian with the following statistics \parencite[e.g.,][]{sanz-alonso_inverse_2023}:
\begin{equation}
  \vec{x}|\vec{y} \sim \mathcal{N}(\vec{m}, \mat{C}), \quad \text{with} \quad \mat{C} = (\hat{\mat{C}}^{-1} + \mat{N}^{-1})^{-1}  \quad \text{and} \quad \vec{m} = \mat{C} \mat{N}^{-1} \vec{y}.
\end{equation}
The best posterior estimate for the balanced signal is therefore~$\vec{m}$, which we reshape into a map for display.

\begin{figure}[t]
  \includegraphics[scale=0.61]{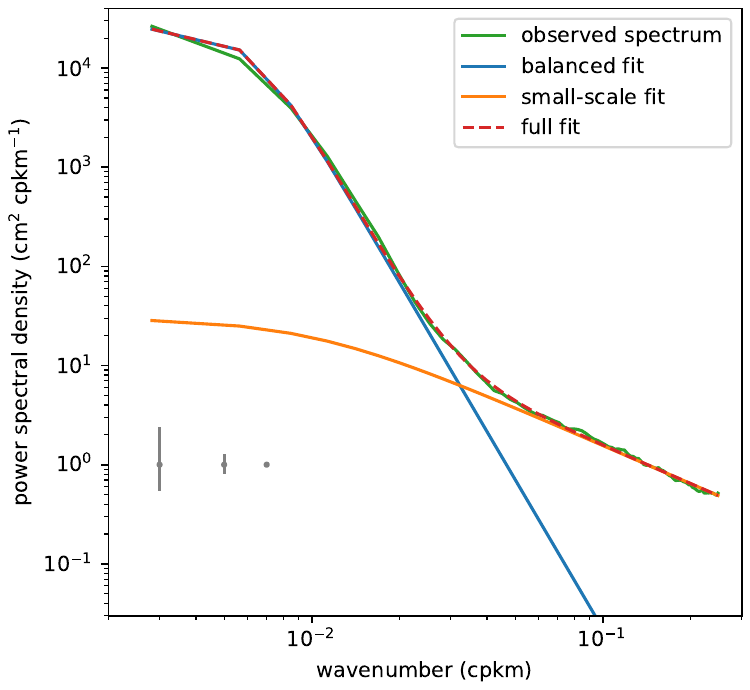}
  \caption{Estimating parameters of the covariance function used in the extraction of the balanced signal in the South Pacific region. Shown is the SSH variance spectrum estimated from a \qty[number-unit-product=-]{350}{\kilo\meter}-long segment in the Antarctic Circumpolar Current (Fig.~\ref{sshmap}). The model~\eqref{slopefiteq} with fixed $s_\mathrm{b} = 5$, $\lambda_\mathrm{s} = \qty{100}{\kilo\meter}/2\pi$, and best-fit parameters is also shown. The vertical gray lines indicate 95\%~confidence intervals assuming that an independent sample is obtained every 10~cycles (left), every cycle (middle), and at every cross-track position (right).}
  \label{maternfit}
\end{figure}
  
We assume stationary and isotropic statistics and prescribe covariance functions based on the spectra estimated from the data. We set $\hat{C}_{ij} = c(r_{ij})$ and $N_{ij} = n(r_{ij})$, where $r_{ij}$ is the distance between the pixels $i$ and~$j$ (assuming Cartesian geometry), and we estimate the covariance functions~$c(r)$ and $n(r)$ as Fourier transforms of fits to the estimated spectrum using the Wiener--Khinchin theorem. We use the same fit as in Section~\ref{results:globalspectra}, except that we now fix the balanced slope at~$s_\mathrm{b} = 5$ to facilitate an analytical Fourier transform, which yields a Meijer-G function for $c(r)$. This slope value is close to that measured using the \qty[number-unit-product=-]{1000}{\kilo\meter}-long segment ($s_\mathrm{b} = \num{4.69+-0.003}$). For the small-scale Mat\'ern model in \eqref{slopefiteq}, the covariance function is
\begin{equation}
  n(r) = \frac{\sqrt{\pi}}{\Gamma(\tfrac{s_\mathrm{s}}{2})} \frac{A_\mathrm{s}}{\lambda_\mathrm{s}} \left( \frac{r}{2\lambda_\mathrm{s}} \right)^\nu K_\nu (\tfrac{r}{\lambda_\mathrm{s}}), \quad \text{where} \quad \nu = \frac{1}{2} (s_\mathrm{s} - 1)
\end{equation}
and $K_\nu$ is a modified Bessel function of the second kind. We again fix $\lambda_\mathrm{s} = \qty{100}{\kilo\meter}/2\pi$ in the fit because is it poorly constrained by the data. We estimate the remaining parameters $\vec{\theta} = (A_\mathrm{b}, \lambda_\mathrm{b}, A_\mathrm{s}, s_\mathrm{s})$ from the fit. We use a \qty[number-unit-product=-]{350}{\kilo\meter}-long segment for this calculation to render the stationarity assumption reasonable. The fit is applied to the spectral estimate from the same segment yielding $A_\mathrm{b} = \qty{2.5e4}{\centi\meter\squared\per\cpkm}$, $\lambda_\mathrm{b} = \qty{163}{\kilo\meter}/2\pi$, $A_\mathrm{s} = \qty{30}{\centi\meter\squared\per\cpkm}$, and $s_\mathrm{s} = \num{1.3}$ (Fig.~\ref{maternfit}). The quality of the fit suggests that the covariance models are reasonable.

\begin{figure}
  \includegraphics[scale=0.61]{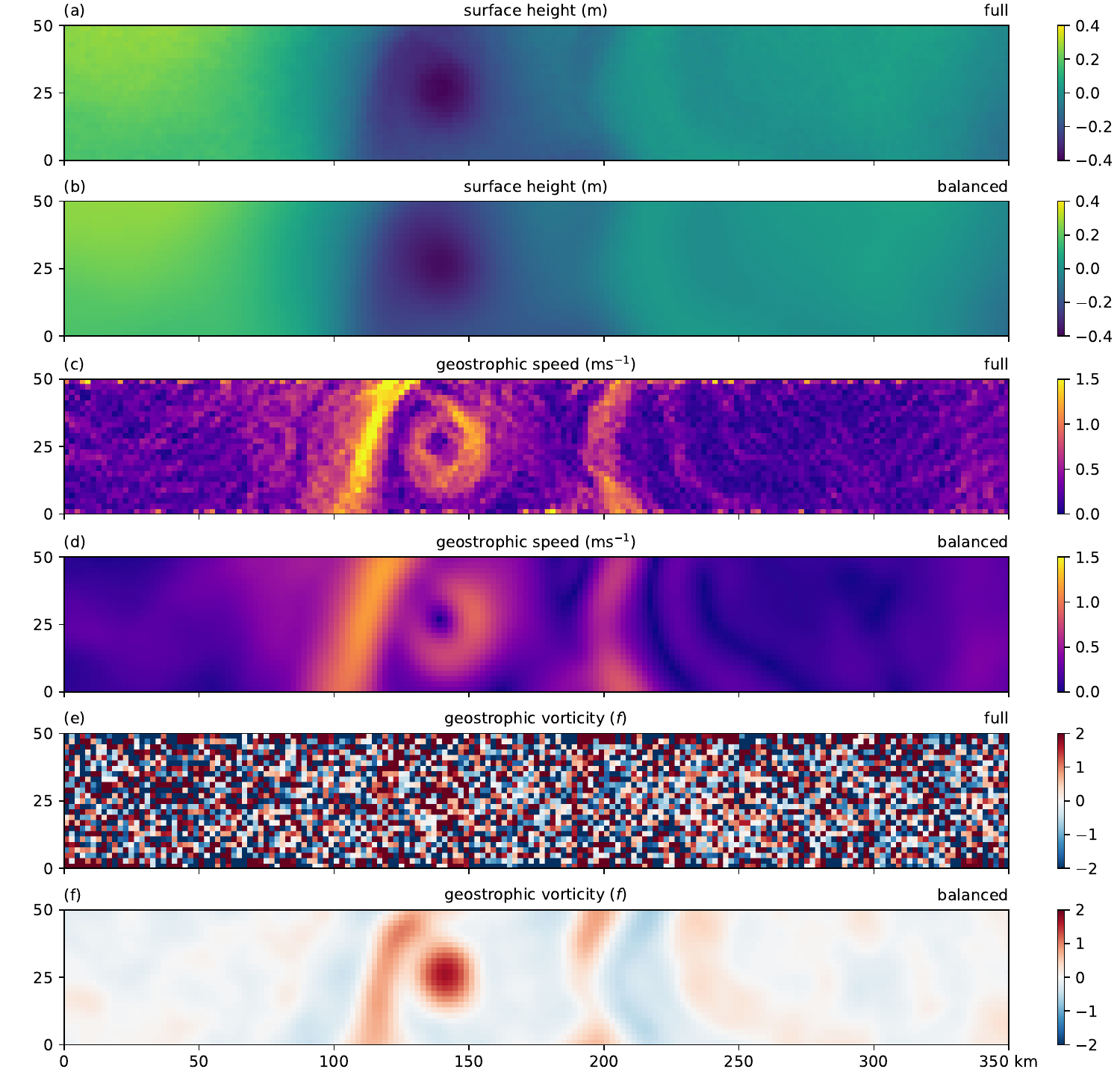}
  \caption{Full and balanced KaRIn signals (right swath) in the South Pacific region on 24-04-2023 (cycle 500). (a,b)~Full and balanced SSH anomalies. (c,d)~Geostrophic speed calculated from the full and balanced signals. (e,f)~Geostrophic vorticity calculated from the full and balanced signals. The geostrophic vorticity is normalized by the planetary vorticity~$f$. Along-track (abscissa) and cross-track (ordinate) distances are given in kilometers and referenced to the northwest corner of the segment (bottom-left corner here; cf., Fig.~\ref{sshmap}).}
  \label{extraction}
\end{figure}

Using these parameters to calculate the covariance matrices $\hat{\mat{C}}$ and~$\mat{N}$ and applying the inversion to the signal observed on KaRIn's right swath on 2023-04-24 (cycle~500), we obtain an estimate~$\vec{m}$ for the balanced signal (Fig.~\ref{extraction}a,b). Because most of the variance is in the balanced signal, the original field~$\vec{y}$ and estimate~$\vec{m}$ look similar, yet close inspection reveals that the inferred balanced signal is smoother, as expected. These differences become more readily apparent as we calculate geostrophic velocities and the vertical component of the geostrophic vorticity:
\begin{equation}
  u_\mathrm{g} = -\frac{1}{f} \pp{\phi}{y}, \qquad v_\mathrm{g} = \frac{1}{f} \pp{\phi}{x}, \qquad \zeta_\mathrm{g} = \pp{v_\mathrm{g}}{x} - \pp{u_\mathrm{g}}{y} + \frac{1}{f} \pp{(u_\mathrm{g},v_\mathrm{g})}{(x,y)},
  \label{sgvort}
\end{equation}
where $\phi = g h$ is the geopotential calculated from the SSH anomaly~$h$. The third term in the geostrophic vorticity is included based on semi-geostrophic theory~\parencite{hoskins_geostrophic_1975}; it makes no qualitative difference in the results. All derivatives are calculated using standard second-order-accurate centered finite differences except near the edges of the swath, where a one-sided second-order-accurate difference scheme is used. The geostrophic flow is strong along the two major fronts at about \qtylist{120;200}{\kilo\meter} along-track position and around the compact cyclone with a radius of some \qty{10}{\kilo\meter} at the center of the swath at about \qty{140}{\kilo\meter} along-track position (Fig.~\ref{extraction}c,d).\footnote{The Fourier transform of a Gaussian with a width of $R = \qty{10}{\kilo\meter}$ is a Gaussian with a width $R^{-1}$ in wavenumber space. The wavenumber spectrum of an isolated vortex with this radius therefore starts falling off exponentially at a wavenumber $k \sim R^{-1}$, which corresponds to a wavelength $\ell = 2\pi/k \sim 2\pi R = \qty{63}{\kilo\meter}$. Most of the variance of such a vortex thus falls into the part of the spectrum dominated by the balanced signal (Fig.~\ref{maternfit}), which is why it is preserved in our extraction.} Smaller-scale structure in the geostrophic speed is removed by the extraction of the balanced signal. The extraction also appears to smear out the fronts and cyclone somewhat. We suspect that capturing such sharp apparently balanced signals would require the prescription of a non-Gaussian prior, which would require more complicated statistical modeling and is not pursued here. Despite this smoothing, the geostrophic vorticities are of the same order as the planetary vorticity, with larger magnitudes in cyclonic regions than anticyclonic regions (Fig.~\ref{extraction}f). This asymmetry between cyclonic and anticyclonic regions is expected from semi-geostrophic theory \parencite{hoskins_atmospheric_1972,thomas_submesoscale_2008} and has been seen in shipboard and airborne current measurements \parencite{rudnick_skewness_2001,shcherbina_statistics_2013,buckingham_seasonality_2016,torres_airborne_2024}. Calculating the geostrophic vorticity from the full signal gives a very noisy field because the two derivatives amplify the small-scale signal (Fig.~\ref{extraction}e).

The observed geostrophic vorticities with magnitudes comparable to the planetary vorticity indicate that an interpretation of these signals using quasi-geostrophic theory is problematic. In the compact cyclone, for example, the balance will be closer to a gradient-wind balance than geostrophic balance. Such inaccuracies will affect estimates of the scale transfer of KE, for example, which are typically based on quasi-geostrophic dynamics \parencite{scott_direct_2005,tulloch_scales_2011}. Semi-geostrophic theory \parencite{hoskins_geostrophic_1975} may offer an attractive alternative. While ageostrophic advection of geostrophic momentum is taken into account, the geostrophic velocities that can be readily computed from the height data remain an element of the theory, and many calculations, when performed in transformed geostrophic coordinates, only involve these geostrophic velocities. The dynamical balance for an axisymmetric vortex, for example, is reasonably approximated by semi-geostrophic theory even when it falls outside of its formal range of applicability \parencite[see the appendix of][]{hoskins_geostrophic_1975}.

\begin{figure}[p]
  \centering
  \includegraphics[scale=0.61]{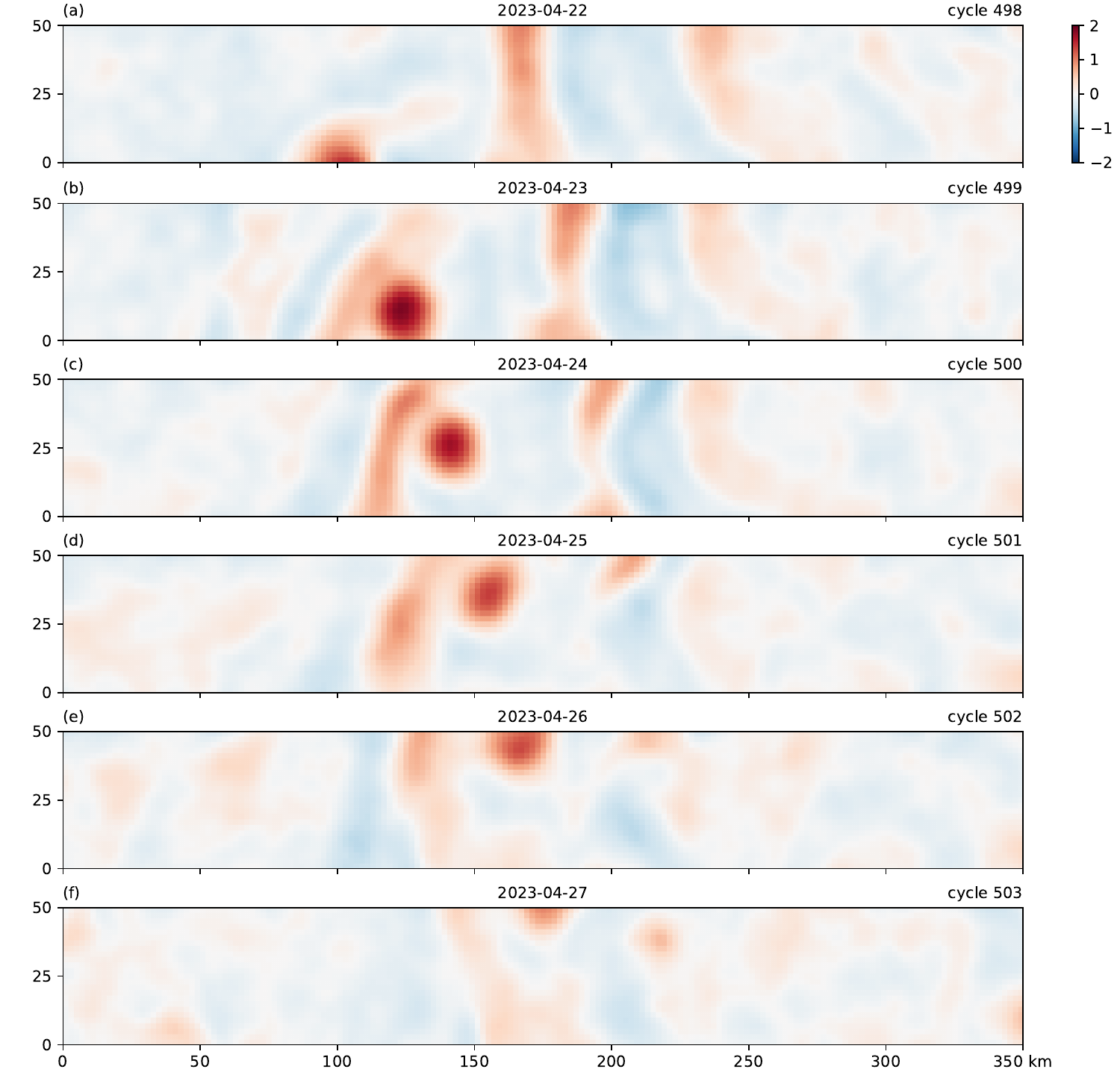}
  \caption{Evolution of the geostrophic vorticity calculated from the balanced signal in the South Pacific region from 2023-04-22 to 2023-04-27 (cycles 489 to~503). All fields are normalized by the planetary vorticity~$f$. Along-track (abscissa) and cross-track (ordinate) distances are given in kilometers and referenced to the northwest corner of the segment (bottom-left corner here; cf., Fig.~\ref{sshmap}).}
  \label{vorticity}
\end{figure}
  
The 1-day resolution of the rapid-repeat phase presents the opportunity to study the evolution of these fronts and vortices. We extract the balanced signal and calculate the geostrophic vorticity as described above for the data from 2023-04-22 to 2023-04-27 (cycles 498 to~503). There is substantial evolution from day to day (Fig.~\ref{vorticity}). The front seen around \qty{200}{\kilo\meter} along-track distance on 2023-04-24 appears to be in the process of breaking up, with the northeast portion rolling up into the compact cyclone centered at \qty{220}{\kilo\meter} along-track distance and \qty{10}{\kilo\meter} cross-track distance on 2023-04-27. The strong cyclone seen in the center of the swath on 2023-04-24 appears to be detaching from the other front and weakening somewhat as it moves eastward over the course of this time period.

\begin{figure}[t]
  \includegraphics[scale=0.61]{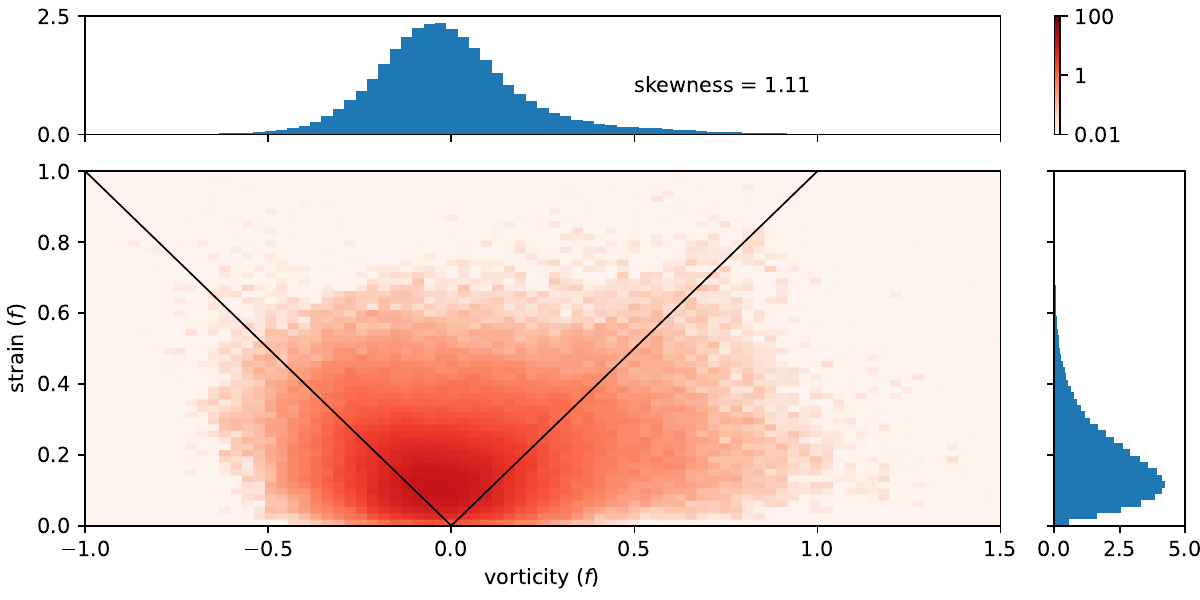}
  \caption{Joint histogram of the geostrophic vorticity and strain rate in the South Pacific region for the rapid-repeat phase. Marginal distributions for vorticity (top) and strain (right) are also shown. All values are normalized by the planetary vorticity~$f$. The black diagonal lines are one-to-one lines for reference.}
  \label{skewness}
\end{figure}

The dominance of strong cyclonic vorticities in this period in late April is not an isolated occurrence. The full-record distribution of geostrophic vorticity, calculated again as described above, has a skewness of~\num{1.11} (Fig.~\ref{skewness}). This skewness is somewhat smaller than that observed in intense submesoscale turbulence \parencite{shcherbina_statistics_2013,buckingham_seasonality_2016}. In similar contrast to previous observations \parencite{shcherbina_statistics_2013}, the geostrophic vorticity's joint distribution with the strain rate
\begin{equation}
  \alpha_\mathrm{g} = \sqrt{\left( \pp{u_\mathrm{g}}{x} - \pp{v_\mathrm{g}}{y} \right)^2 + \left( \pp{v_\mathrm{g}}{x} + \pp{u_\mathrm{g}}{y} \right)^2}
\end{equation}
shows little concentration along the one-to-one line (Fig.~\ref{skewness}). This indicates a decreased prevalence of fronts, in which regions of strong vorticity tend to align with regions of strong strain. These differences with previous observations could be the result of the in situ observations capturing somewhat smaller scales (order \qty{1}{\kilo\meter}) than the balanced signal from KaRIn \parencite[cf., Fig.~4 in][]{chelton_prospects_2019}, the submesoscale turbulence not being as energetic in this South Pacific region and at this time of year than in the wintertime North Atlantic, or a combination of the two. This is worth exploring further with the longer record available during the SWOT science phase.

\section{Conclusions}
\label{conclusions}

SWOT's wide-swath KaRIn data capture at least four different signals: a balanced signal dominating at mesoscales as well as submesoscales in high-energy regions (western boundary current extensions and the Southern Ocean), low-mode internal tides contributing substantially to the submesoscale signal outside of high-energy regions, a small-scale signal that appears to be linked to surface gravity waves, and small-scale geoid variations caused by bathymetry.

The spectral character of the balanced signal broadly conforms with expectations from in situ observations and numerical simulations. The SSH variance spectrum falls off steeply with wavenumber, with most regions exhibiting power laws between $k^{-4}$ and $k^{-6}$. Isolating these balanced signals in the two-dimensional data from KaRIn reveals strong cyclones and sharp fronts in an example region in the South Pacific. The geostrophic relative vorticity frequently reaches magnitudes comparable to the planetary vorticity~$f$, challenging the usual interpretation of balanced altimetry signals using geostrophic balance and quasi-geostrophic theory. Semi-geostrophic theory may offer an alternative.

The rapid-repeat phase assessed here is too short to robustly distinguish between energetic and weak submesoscale turbulence in the balanced signals and to study seasonal variations in this turbulence and its interaction with the mesoscale. This should be feasible with the science phase data once enough cycles have accumulated to robustly remove small-scale geoid variations.

Away from the major current systems, SSH variance spectra typically exhibit peaks corresponding to low-mode baroclinic tides. No attempt has been made here to study or remove these signals.

A small-scale signal with a much flatter SSH variance spectrum is pervasive in the KaRIn data. Its characteristics are at odds with expectations for the internal-wave continuum: the spectral slope is too small, it is much stronger in the southern than the northern hemisphere, and its amplitude has a seasonal trend that is the opposite of what is expected for internal waves. Instead, the amplitude of this signal is strongly correlated in space and time with the heights of surface gravity waves, suggesting aliased surface waves as a leading contender for this signal. More work is needed to better understand what sets the character of this signal and exclude other possibilities, especially other processes that are also directly forced by the winds.

\section*{Open research section}

\sloppy

The SWOT mission data were taken from the NASA Physical Oceanography Distributed Active Archive Center (PODAAC) \parencite{SWOT2024}, available at \url{https://doi.org/10.5067/SWOT-SSH-2.0}. The significant wave height is computed with data from ERA5 reanalysis \parencite{hersbach_era5_2020}. The geoid model is taken from \url{https://doi.org/10.25921/fd45-gt74}. The DUACS SSHA is taken from \url{https://doi.org/10.48670/moi-00149}. The ECCO~v4 data used to compute tidal wavenumber is available at \url{ https://doi.org/10.5281/zenodo.4533349}. The code used for the data analysis can be found at \url{https://doi.org/10.5281/zenodo.16693002}.
\fussy

\section*{Acknowledgments}

This work was supported by NASA grants 80NSSC20K1140 and 80NSSC24K1652. The authors gratefully acknowledge stimulating discussions with Albion Lawrence and Jack Skinner as well as feedback on the manuscript from Dudley Chelton, Carl Wunsch, and two anonymous reviewers.
\end{document}